\documentclass[journal]{IEEEtran}
\usepackage[cmex10]{amsmath}
\usepackage{amsfonts,amssymb}
\usepackage{epsfig,subfigure}
\usepackage{verbatim}
\usepackage{multirow}

\newtheorem{thm}{Theorem}
\newtheorem{lem}{Lemma}
\newtheorem{cor}{Corollary}

\begin{document}

\title{Unconditionally Secure Computation on Large Distributed Databases with Vanishing Cost}

\author{Ye~Wang, Shantanu~Rane, Prakash~Ishwar, Wei~Sun
\thanks{Y~.Wang and S.~Rane are with Mitsubishi Electric Research Laboratories, Cambridge, MA 02139 USA. E-mail: {\tt \{yewang,rane\}@merl.com}.
Portions of this work was conducted while Y. Wang was with the Department of Electrical and Computer Engineering, Boston University, Boston, MA 02215.}
\thanks{P.~Ishwar is with the Department of Electrical and Computer Engineering, Boston Unversity, Boston, MA 02215 USA. E-mail: {\tt pi@bu.edu}.}
\thanks{W.~Sun is with Symanta Inc., Kitchener, Ontario N2G1H6 Canada. E-mail: {\tt sunwei\_hk@yahoo.com}.}
\thanks{The third author acknowledges support from the US NSF under award number \#0915389.
The views and conclusions contained in this article are those of the authors and should not be interpreted as necessarily representing the official policies, either expressed or implied, of the US NSF or MERL.}
\thanks{Portions of this work previously appeared in a conference paper \cite{WangRSI-Allerton10-SFCwVCommCost}.}
}

\maketitle

\begin{abstract}
Consider a network of $k$ parties, each holding a long sequence of $n$
entries (a database), with minimum vertex-cut greater than $t$. We
show that any empirical statistic across the network of databases can
be computed by each party with perfect privacy, against any set of $t
< k/2$ passively colluding parties, such that the worst-case distortion and
communication cost (in bits per database entry) both go to zero
as $n$, the number of entries in the databases, goes to infinity. This
is based on combining a striking dimensionality reduction result for
random sampling with unconditionally secure multi-party computation protocols.
\end{abstract}


\section{Introduction}

We consider a secure multi-party computation problem
involving a connected network of $k$ parties, where the $i^{\mathrm{th}}$ party
($i = 1,\ldots,k$) has a deterministic sequence $\mathbf{x}_i :=
x_{i,1},\ldots,x_{i,n}$ of $n$ entries (e.g., a database), and wishes
to compute a function of the form $f_i(\mathbf{x}^k) = \frac{1}{n}
\sum_{l=1}^n \phi_i(x_{1,l},\ldots,x_{k,l})$, where $\mathbf{x}^k :=
(\mathbf{x}_1,\ldots,\mathbf{x}_k)$ denotes the sequences of all the
parties. Functions of this form arise in the context of computing
empirical statistics of observations and are referred to as normalized
sum-type functions. The objective is to construct a (possibly
randomized) message-passing protocol that securely computes these
functions with high accuracy and low communication cost.  We assume
that the parties are semi-honest (passive), which means that they will
correctly follow the steps of the protocol, but will attempt to infer
the maximum possible information about each other's sequences from the
data available to them.  We require unconditional privacy against
coalitions of size $t < k/2$.  This means (in a strong statistical
sense) that any coalition is unable to infer any information about the
other parties' sequences other than what can be inferred from their
own sequences and function estimates.

In contrast to other secure multi-party computation formulations such
as~\cite{BenOrGwW-ACM88-CTNCFTDC}, \cite{ChaumCrepDam-ACM88-MPUSP},
\cite{CrepeauSSW-Eurocrypt06-ITSecCond2PSFE}, \cite{WangIshwarISIT09},
which aim to make the probability of error of each function estimate,
$\Pr\big[\hat{F}_i(\mathbf{x}^k) \neq f_i(\mathbf{x}^k)\big]$, where
$\hat{F}_i(\mathbf{x}^k)$ is the function estimate of the $i^{\mathrm{th}}$
party, equal to zero or negligible, we consider a novel
distortion-theoretic approach that aims to minimize the maximum
expected absolute error of each function estimate,
\[
\max_{\mathbf{x}^k} E\left[\big|\hat{F}_i(\mathbf{x}^k) - f_i(\mathbf{x}^k)\big|\right], \quad i \in \{1, \ldots, k\},
\]
where the expectation is with respect to any randomness inherent to
the protocol in generating the estimate $\hat{F}_i(\mathbf{x}^k)$.
The distortion is the worst-case expected absolute error across all
deterministic sequences $(\mathbf{x}^k)$.  The communication cost is
given by the number of bits of transmission required by a protocol
divided by $n$, the length of the sequences, i.e., the total number of
bits per sample or bitrate.  Our main result is the construction of
unconditionally private protocols that compute any normalized sum-type
functions with both vanishing distortion and vanishing communication
cost as $n \rightarrow \infty$, in networks whose minimum
vertex-cut\footnote{The minimum number of vertices (parties) whose
  removal leaves the network disconnected.} is greater than $t$.
While the (perfectly) secure multi-party computation techniques
of~\cite{BenOrGwW-ACM88-CTNCFTDC} can be used to compute any
normalized sum-type function without error (zero distortion), they
require $O(n)$ transmissions (see Section~\ref{sec:PolyProtocols}) and
hence have non-vanishing communication cost.

The key to our result is the realization that any normalized sum-type
function can be evaluated accurately even after drastically reducing
the dimensionality of its inputs.  Although there are elegant
dimensionality reduction results which show that distances can be
approximately preserved by mapping high-dimensional signals into a
low-dimensional subspace~\cite{IndykMatousek-04-LDEoFMS},
\cite{JohnsonLindenstrauss-CoM85-ELmiHS}, they are within a {\em
  centralized} computation context and without privacy constraints.
We consider a much more basic dimensionality reduction that is
achieved by a simple random sampling. It was shown
in~\cite{AhlswedeZhang-LNCS06-EstDist} that an accurate estimate of
the joint type of two sequences can be produced from a
randomized sampling of the sequences.  We develop a
much simpler and more general analysis of randomized sampling,
which allows us to analyze the expected distortion, and apply this
result to create accurate estimates of normalized sum-type functions
in a manner that is both secure and communication-efficient.  The
randomization in the sampling is crucial for achieving vanishing
distortion with a worst-case distortion criterion. Sampling by a
factor much smaller than $n$ allows us to use fewer invocations of the
secure computation primitives of~\cite{BenOrGwW-ACM88-CTNCFTDC} while
securely producing the function estimate with vanishing communication
cost.

It is important to highlight the distinction between a vanishing
error-probability criterion and a vanishing expected-distortion
criterion.  Distributed computation of the joint type with a vanishing
error-probability needs a strictly positive communication bitrate
which does not vanish with increasing
blocklength~\cite{AhlswedeZhang-LNCS06-EstDist},
\cite{AhlswedeC-IT1981-OneBitHardAsFullInfo},
\cite{ElGamal-IT1983-SimpleProofOfACOneBit}, whereas the bitrate
vanishes with blocklength for vanishing
expected-distortion~\cite{AhlswedeZhang-LNCS06-EstDist}.  Our
distortion-theoretic approach to secure multi-party computation thus
trades exact computation for arbitrarily high accuracy in order to
gain the advantage of vanishing communication cost.  This makes these
results particularly relevant to applications such as secure
statistical analysis of distributed databases, where data size is
overwhelming and only a highly accurate, but not exact, computation is
necessary.

\section{Problem Formulation}
\label{sec:Formulation}

There are $k$ parties, where each party $i \in \{1,\ldots,k\}$ has
private data modeled as a deterministic sequence of $n$ symbols,
$\mathbf{x}_i := (x_{i,1}, \ldots, x_{i,n}) \in \mathcal{X}_i^n$, from
some finite alphabet $\mathcal{X}_i$.  Each party $i \in
\{1,\ldots,k\}$ wishes to compute a function, $f_i(\mathbf{x}^k)$, of
all of the sequences $\mathbf{x}^k :=
\mathbf{x}_1,\ldots,\mathbf{x}_k$.  The objective is to design a
multi-party protocol that allows all of the parties to securely
compute their desired functions with high accuracy and low
communication cost.  In the remainder of this section, we describe the
class of functions of interest and make precise the notions of
accuracy, security, and communication cost.

The class of functions that we consider are the rational-valued, {\em
normalized sum-type functions}, which means that for each $i \in
\{1, \ldots, k\}$, the function $f_i : \mathcal{X}_1^n \times \ldots
\times \mathcal{X}_k^n \rightarrow \mathbb{Q}$ is of the form
\begin{eqnarray} \label{eqn:SumTypeFunc}
f_i(\mathbf{x}^k) = \frac{1}{n} \sum_{l=1}^n \phi_i(x_{1,l},\ldots,x_{k,l}),
\end{eqnarray}
for some function $\phi_i:\mathcal{X}_1 \times \ldots \times
\mathcal{X}_k \rightarrow \mathbb{Q}$.

A protocol is a sequence of instructions that specifies how the parties
will interact.  The execution of a protocol consists of a sequence of
local computations and message transfers between the parties via
bi-directional, error-free channels that are available between pairs
of parties.  The messages sent at any stage of the execution of the
protocol may depend on previously received messages, the sequences
that are available to the parties sending the messages, and any
independent local randomness that is generated.  At the end of the
protocol, each party produces a function estimate
$\hat{F}_i(\mathbf{x}^k)$.  While the inputs $\mathbf{x}^k$ and the
functions $f_i$ are deterministic, the estimates
$\hat{F}_i(\mathbf{x}^k)$ may be random due to randomness used by the
protocol.  We define the {\em view} of a party as the set of all
messages sent or received, and any local randomness generated by that
party during the execution of the protocol.  Let the random variables
$V_1, \ldots, V_k$ denote the views of each of the $k$ parties after
the execution of the protocol.

\emph{Accuracy:}
The distortion criterion to be minimized is the {\em maximum expected
  absolute error}:
\begin{eqnarray*}
e_i(n) := \max_{\mathbf{x}^k} E\left[\big|\hat{F}_i(\mathbf{x}^k) -
  f_i(\mathbf{x}^k)\big|\right], \quad i \in \{1, \ldots, k\},
\end{eqnarray*}
where the expectation is with respect to the local randomness that is
generated in the execution of the protocol.  We emphasize that
$\mathbf{x}^k$ are deterministic sequences and the distortion is the
worst-case expected absolute error over all possible sequences.

\emph{Privacy:}
Since an approximation protocol computes an estimate instead of the
exact function we must require these two privacy conditions:
\begin{enumerate}
\item \emph{Protocol Privacy:} The protocol should only reveal the approximate computation.
\item \emph{Approximation Privacy:} The approximate computation should not inherently reveal more information than the exact computation would have.
\end{enumerate}
The first privacy condition is a property of the protocol, while the
second is a property of the approximation that it produces. In the
next two paragraphs we formalize this first condition, requiring that
the protocols securely compute the randomized function estimate with
{\em unconditional protocol privacy}. Later on, in Section~\ref{sec:Privacy},
we will analyze the privacy implications inherent to replacing an
exact computation with the approximate estimate that our proposed
protocols provide, and propose an approximation privacy condition
that is satisfied if the approximation can be made arbitrarily close
to the exact computation with arbitrarily high probability.

We will consider protocols that are unconditionally private
against a colluding coalition of $t$ {\em semi-honest} parties $T
\subset \{1, \ldots, n\}$, for $t < k/2$.  The semi-honest assumption
requires that each party will correctly follow the protocol, but may attempt to gather information about other parties' data from the view available to it.
Unconditional privacy requires that after the execution of the
protocol, the views of any coalition of size $t < k/2$ do not reveal
any more information, in a strong statistical sense, about the private
inputs and function estimates of the other parties, than what can be
inferred from the function estimates and inputs of the coalition.  We
use the notation $V_T := \{V_i : i \in T\}$ and $\mathbf{x}_T :=
\{\mathbf{x}_i : i \in T\}$ to denote the views and input sequences of
the parties in coalition $T$, and likewise, $\mathbf{x}_{\bar{T}} :=
\{\mathbf{x}_i : i \in \bar{T}\}$ to denote the input sequences of the
parties {\em not} in coalition $T$.

A protocol is {\em unconditionally private} if for all coalitions $T
\subset \{1, \ldots, k\}$, with $|T| < k/2$, the {\em distribution} of
the views of the coalition conditioned on the function estimates of
the coalition is only parameterized by input sequences of the
coalition and the views of the coalition are independent of the
function estimates of the non-coalition parties, that is, for all
\begin{align*}
(\mathbf{x}_T, \mathbf{x}_{\bar{T}}, \mathbf{x'}_{\bar{T}}) &\in
\Big( \prod_{i \in T} \mathcal{X}_i \Big) \times 
\Big( \prod_{j \in \bar{T}} \mathcal{X}_j \Big) \times 
\Big( \prod_{k \in \bar{T}} \mathcal{X}_k \Big), \\
(\hat{f}_T, \hat{f}_{\bar{T}}) &\in 
\Big( \prod_{i \in T} \mathcal{\hat{F}}_i \Big) \times 
\Big( \prod_{j \in \bar{T}} \mathcal{\hat{F}}_j \Big),
\end{align*}
such that,
\[
P_{\hat{F}_T, \hat{F}_{\bar{T}}} (\hat{f}_T, \hat{f}_{\bar{T}} ; \mathbf{x}_T,\mathbf{x}_{\bar{T}}),
P_{\hat{F}_T, \hat{F}_{\bar{T}}} (\hat{f}_T, \hat{f}_{\bar{T}} ; \mathbf{x}_T,\mathbf{x'}_{\bar{T}}) > 0,
\]
we have that
\[
P_{V_T|\hat{F}_T, \hat{F}_{\bar{T}}} (v_T | \hat{f}_T, \hat{f}_{\bar{T}} ; \mathbf{x}_T,\mathbf{x}_{\bar{T}}) = P_{V_T|\hat{F}_T}(v_T|\hat{f}_T;\mathbf{x}_T,\mathbf{x'}_{\bar{T}}).
\]

This privacy condition is based on a rather strong notion of
statistical indistinguishability. Note that for an unconditionally
private protocol, if the deterministic input sequences $(\mathbf{x}_1,
\ldots, \mathbf{x}_k)$ were replaced with sequences of random
variables $(\mathbf{X}_1, \ldots, \mathbf{X}_k)$ drawn from {\em any
  distribution}, then the views of any coalition $T \subset \{1,
\ldots, k\}$, with $|T| < k/2$, would satisfy the following Markov
chain relationship,
\begin{eqnarray*}
V_T - (\mathbf{X}_T, \hat{F}_T) - (\mathbf{X}_{\bar{T}}, \hat{F}_{\bar{T}}).
\end{eqnarray*}
These Markov chain conditions are analogous to the two-party
conditional mutual information conditions
of~\cite{CrepeauSSW-Eurocrypt06-ITSecCond2PSFE,WangIshwarISIT09} when
appropriately adapted to our problem involving multiple semi-honest
parties.

\emph{Communication Cost:}
For a given protocol, let $r$ denote the number of bits necessary to
send all of the messages required by the protocol.  The communication
cost of a protocol is given by the {\em rate} $R := (r/n)$. 

\begin{figure}[thpb]
\centering
\subfigure[Fully connected]{ \includegraphics[width=1.5in]{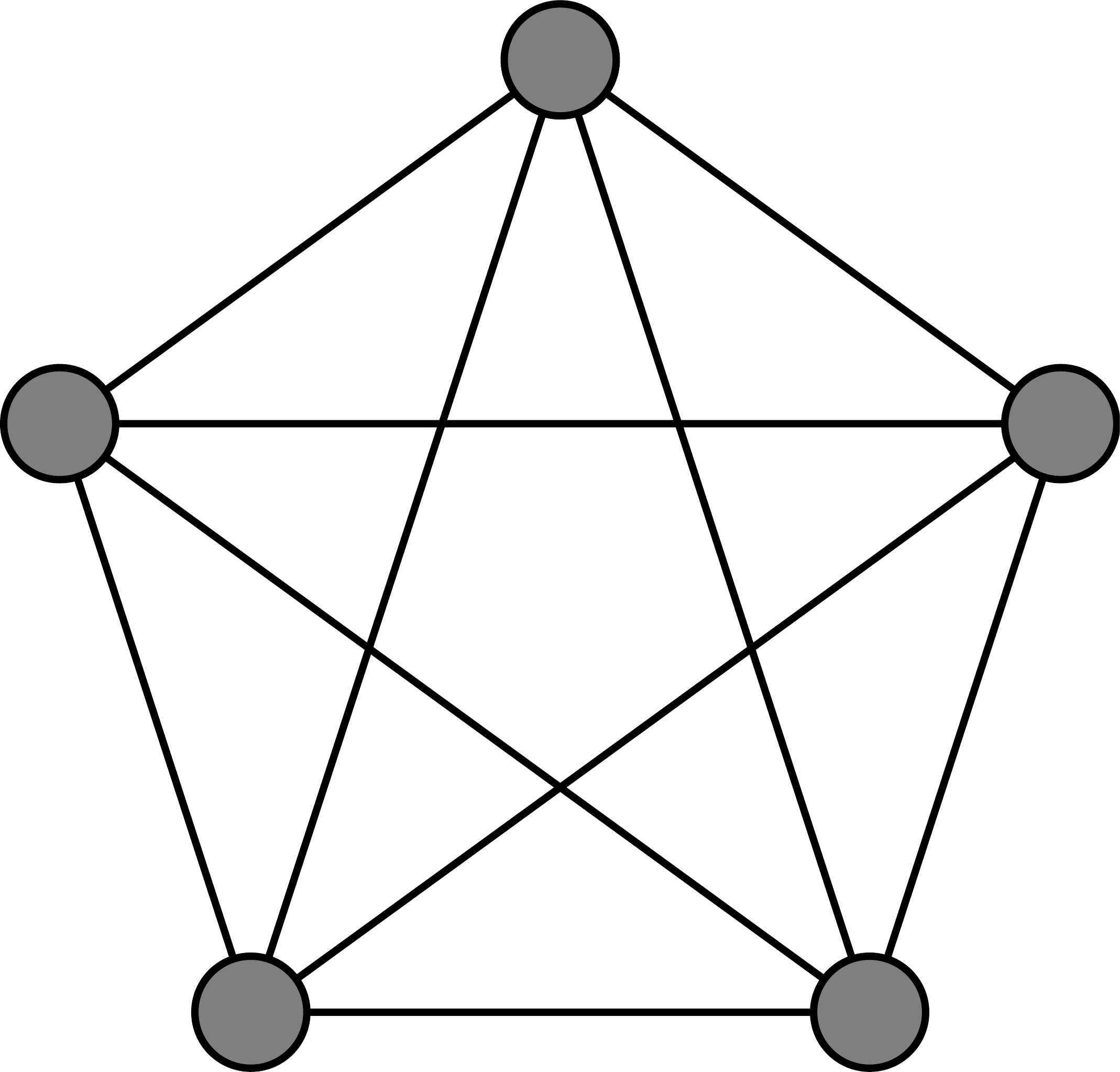} }
\hspace{0.25in}
\subfigure[Partially connected]{ \includegraphics[width=1.5in]{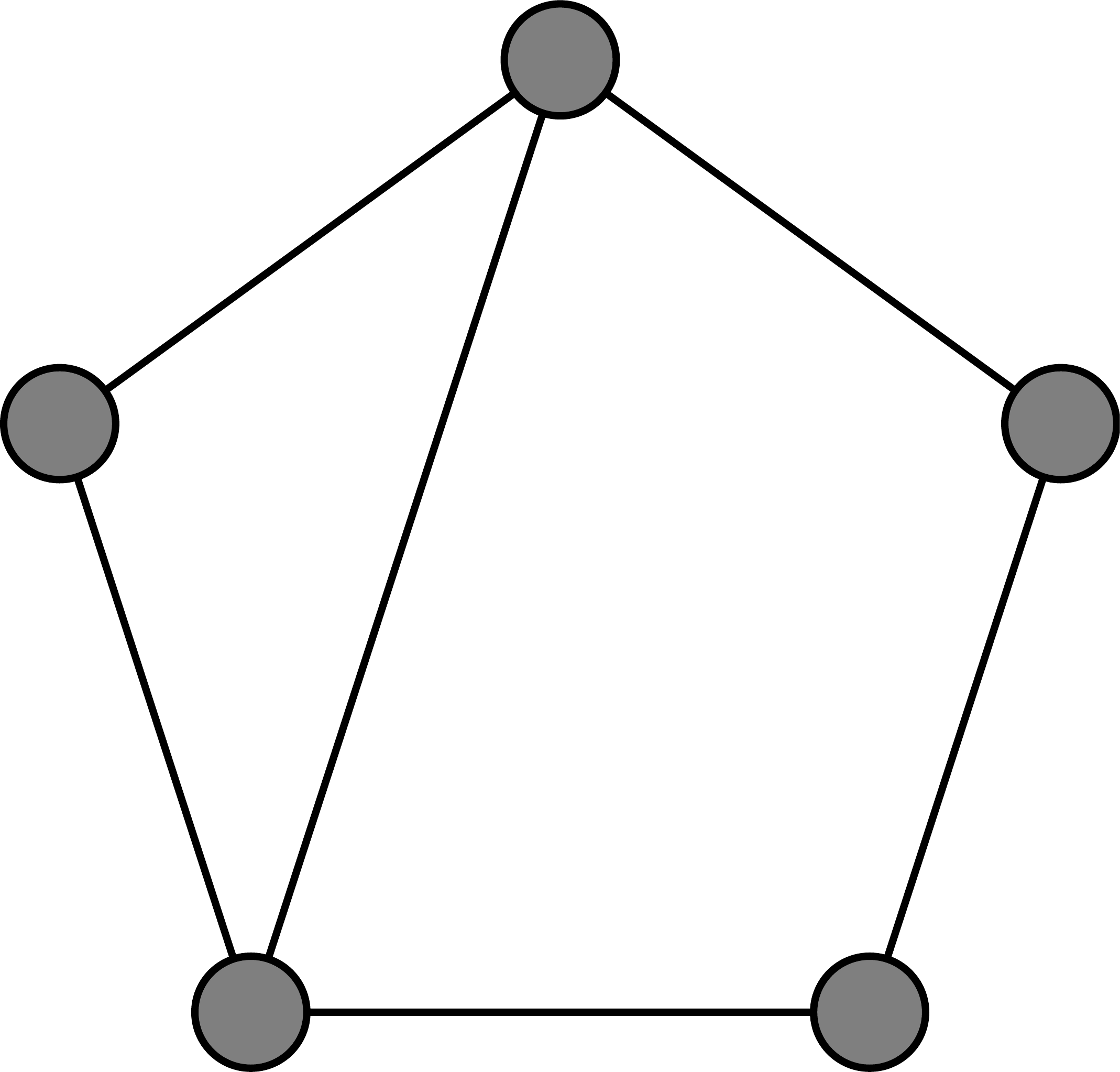} }
\caption{(a) A fully connected network of $5$ parties. Each pair of
  parties may communicate over a private error-free channel.  (b) A
  partially connected network of $5$ parties with a minimum vertex-cut
  of $2$. Parties may communicate only over a subset of the channels
  available in the fully connected network. Every pair of parties has
  $2$ non-intersecting paths connecting them.}
\label{fig:NetworkComp}
\end{figure}

We require that the network of parties has a minimum vertex-cut
greater than $t$ which essentially ensures that there are a sufficient
number of non-intersecting paths between any pair of parties
(cf.~Figure~\ref{fig:NetworkComp}) and, as will become clear in the
sequel, makes unconditional privacy attainable. We will, however,
first develop unconditionally private protocols for a {\em fully
  connected network} where a bi-directional error-free channel exists
between every pair of parties. We will then show how these protocols
can be re-purposed for {\em partially connected networks} with minimum
vertex-cut $> t$ where bi-directional error-free channels are
available only for a subset of all pairs of parties. The network
available can be modeled as a complete (fully connected network) or
incomplete (partially connected network) graph as illustrated in
Figure~\ref{fig:NetworkComp}.

\section{Main Results}

Our main result is that as $n \rightarrow \infty$, all normalized
sum-type functions can be computed with arbitrarily high accuracy,
vanishing communication cost, and unconditional privacy.  The
following theorem asserts that we can construct protocols with
properties that enable this claim for fully-connected networks.

\begin{thm} \label{thm:MainThm}
For all parameters $m \in \{1,\ldots,n\}$ and any normalized sum-type
functions $f_1, \ldots, f_k$, there exist unconditionally private
randomized protocols for fully-connected networks, with maximum
expected absolute errors bounded by
\begin{eqnarray} \label{eqn:ThmErrorBound}
e_i \leq \frac{\|\phi_i\|_2}{\sqrt{m}}, \quad \text{for } i \in \{1, \ldots, k\},
\end{eqnarray}
where $\phi_i$ is the sample-wise function in the expansion given by (\ref{eqn:SumTypeFunc}), and total communication cost on the order of
\begin{eqnarray} \label{eqn:ThmRate}
R = \frac{O(m \log n)}{n}.
\end{eqnarray}
\end{thm}

\IEEEproof This result follows from our analysis of a general function
approximation method in Section~\ref{sec:KeyResults} and the protocols
constructed in Section~\ref{sec:Protocols} that securely realize this
approximation. See Appendix~\ref{app:MainProof} for the detailed
proof.  \endproof

An interesting feature of this result is that the error is independent
of the length of the sequences $n$ and is only controlled by the
parameter $m$.  Furthermore, the scaling of rate as a function of $m$
and $n$ leads to the following corollary which states that as the
sequence length $n$ grows, both asymptotically vanishing distortions
and communication costs can be achieved by suitably scaling the
parameter $m$.

\begin{cor} \label{cor:vanishing}
For any normalized sum-type functions $f_1, \ldots, f_k$, there exist
unconditionally private randomized protocols for fully-connected
networks such that as $n$ goes to infinity, the maximum expected
absolute errors $e_i$, for all $i \in \{1,\ldots,k\}$, go to zero and
simultaneously the communication cost $R$ also goes to zero.
\end{cor}

\IEEEproof Choose a sequence of parameters $m_n$ in
Theorem~\ref{thm:MainThm} such that as $n$ goes to infinity,
\begin{eqnarray*}
m_n \rightarrow \infty, \quad \frac{m_n \log n}{n} \rightarrow 0.
\end{eqnarray*}
For example, $m_n = \log(n)$, or $m_n = n^{1-\epsilon}$, for $\epsilon
\in (0,1)$.
\endproof

Thus, in the context of overwhelmingly large databases, we can
leverage this scaling to achieve low distortion and bit rate. Our next
theorem considers the scenario in which the parties can only
communicate through a partially connected network that only provides
channels for a subset of the pairs of parties.

\begin{thm} \label{thm:NetworkThm}
Let the $k$-vertex graph $G$ represent the communication network of
the $k$ parties, where there is an edge between a pair of vertices if
there is a channel available between the corresponding pair of
parties.  If the graph $G$ has minimum vertex cut, i.e., the minimum
number of vertices whose removal leaves $G$ disconnected, greater than
$t$, then the protocols from Section~\ref{sec:Protocols} that satisfy
Theorem~\ref{thm:MainThm} can be re-purposed for use in the given
network, while remaining unconditionally private and achieving the
same error bound as in (\ref{eqn:ThmErrorBound}) and total
communication cost on the same order as in (\ref{eqn:ThmRate}).
\end{thm}

\IEEEproof
See Appendix~\ref{app:NetworkProof} for the detailed proof.
\endproof

The implication of this result is that as long as the network is
sufficiently connected, the performance guarantees of
Theorem~\ref{thm:MainThm} will continue to hold for a partially
connected network while maintaining the same order of magnitude
communication cost.

\section{Joint Type and Function Estimation} \label{sec:KeyResults}

In this section, we provide key results that enable the construction
of protocols that attain the performance guarantees of
Theorem~\ref{thm:MainThm}.  Our protocols, constructed in
Section~\ref{sec:Protocols}, produce a function estimate generated
from only a random sampling of the sequences
$(\mathbf{x}_1,\ldots,\mathbf{x}_k)$.  This technique is a simplified
and generalized version of the striking dimensionality reduction
result that an accurate estimate of the joint type of a pair of
sequences $(\mathbf{x}_1,\mathbf{x}_2)$ can be produced from only a
random sampling of the sequences~\cite{AhlswedeZhang-LNCS06-EstDist}, however, our result has been arrived at by very different means of analysis.
We combine this result with the fact that a normalized sum-type function can be computed from the joint type in order to form an accurate function estimate.
We will first discuss the dimensionality reduction result in
Section~\ref{sec:TypeResult} in order to analyze the function estimate
in Section~\ref{sec:FuncEst}.

\subsection{Estimating the Joint Type} \label{sec:TypeResult}

Outside the context of secure function computation, we first discuss a
dimensionality reduction result concerning the estimation of the joint
type (empirical distribution)
$P_{\mathbf{x}_1,\ldots,\mathbf{x}_k}(x_1,\ldots,x_k)$ from only a
random sampling of the sequences
$(\mathbf{x}_1,\ldots,\mathbf{x}_k)$.  For a given sampling
parameter $m \in \{1,\ldots, n\}$, the random sampling procedure
chooses $m$ locations $L_1,\ldots,L_m \in \{1,\ldots,n\}$ uniformly
{\em without} replacement.  Let $(\mathbf{x}_1,\ldots,\mathbf{x}_k)_L
:= \{(x_{1,l}, \ldots, x_{k,l}), l \in L\}$ denote the sequences
obtained by sampling $(\mathbf{x}_1,\ldots,\mathbf{x}_k)$ according
to the locations in the index set $L := \{ L_1,\ldots,L_m \}$.

The full and partial frequency functions (histograms) $N,M :
\mathcal{X}_1 \times \ldots \times \mathcal{X}_k \rightarrow
\{0,\ldots,n\}$ are defined by
\begin{eqnarray*}
N(x^k) &:=& n P_{\mathbf{x}^k}(x^k), \\
M(x^k) &:=& \big|\{l \in L: (x_{1,l},\ldots,x_{k,l}) = x^k \}\big|.
\end{eqnarray*}
Note that $N(x^k)$ is a deterministic quantity based on the sequences $\mathbf{x}_1,\ldots,\mathbf{x}_k$. However, $M(x^k)$ is a
hypergeometric random variable, since $m$ samples are chosen without
replacement from a set of size $n$, where $N(x^k)$ of them can contribute
to the value of $M(x^k)$.

Let the estimate of the joint type be given by
\begin{equation} \label{eqn:TypeEst}
\hat{P}_{\mathbf{x}^k}(x^k) := \frac{M(x^k)}{m}.
\end{equation}

\begin{lem} \label{lem:TypeProps}
For any finite-alphabet sequences $(\mathbf{x}_1,\ldots,\mathbf{x}_k)
\in \mathcal{X}_1^n \times \ldots \times \mathcal{X}_k^n$, sampling
parameter $m \in \{1, \ldots, n\}$, and $x^k \in \mathcal{X}_1 \times
\ldots \times \mathcal{X}_k$, the randomly sampled type estimate
given by (\ref{eqn:TypeEst}) has the following properties:
\begin{eqnarray*}
E\big[\hat{P}_{\mathbf{x}^k}(x^k)\big] &=& P_{\mathbf{x}^k}(x^k), \\
\mathrm{Var}\big[\hat{P}_{\mathbf{x}^k}(x^k)\big] &=& \frac{N(x^k)(n-N(x^k))(n-m)}{m n^2 (n-1)} \\
&\leq& \frac{N(x^k)}{m n} = \frac{P_{\mathbf{x}^k}(x^k)}{m}, \\
E \left[ \big\|\hat{P}_{\mathbf{x}^k} - P_{\mathbf{x}^k} \big\|_2 \right] &\leq& \frac{1}{\sqrt{m}}.
\end{eqnarray*}
\end{lem}

\IEEEproof
See Appendix~\ref{app:TypeProps} for the detailed proof.
\endproof

Thus, for any set of sequences $(\mathbf{x}_1,\ldots,\mathbf{x}_k)$ of
any length $n$, the estimate of the joint type of the sequences
produced from only $m$ random samples achieves expected $L_2$-error
inversely proportional to $\sqrt{m}$.

\begin{figure}[ht]
\centering
\subfigure[Uniform type over values $\{0, \ldots, 9\}$.]{
\includegraphics[width=3.2in]{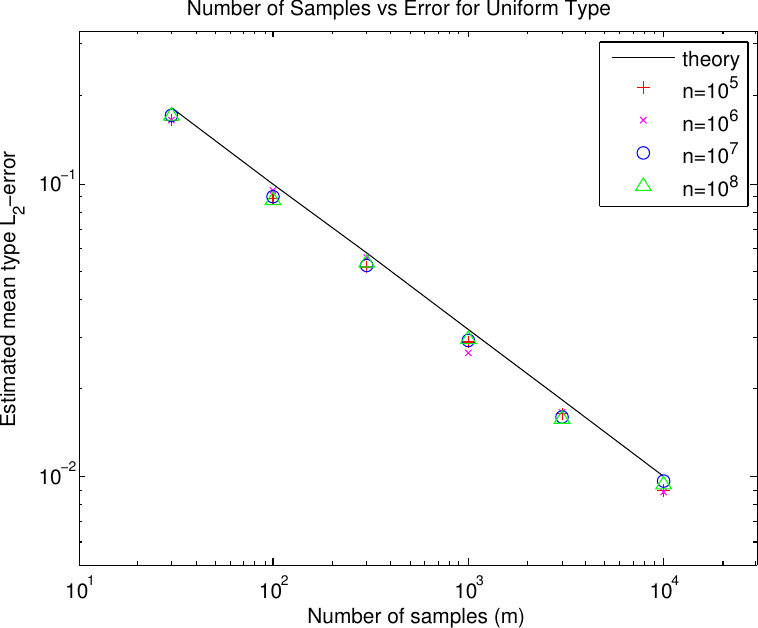}
\label{fig:TypeUnifSubFig}}
\subfigure[Binomial type over values $\{0, \ldots, 9\}$ with mean $4.5$.]{
\includegraphics[width=3.2in]{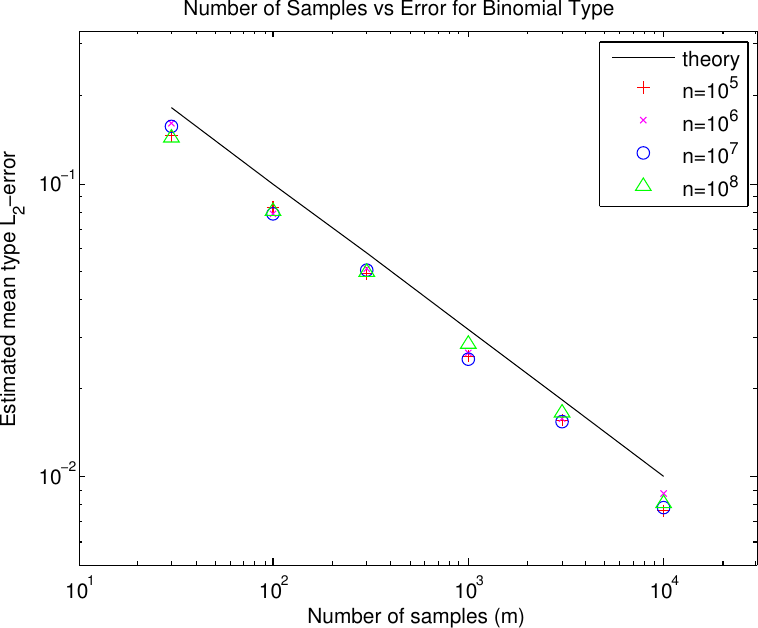}
\label{fig:TypeBinomSubFig}}
\caption{\label{fig:TypeEvaluationExample}
We conducted the simulation across various choices of the parameters $n$ (length of the sequences) and $m$ (number of samples), and for both a uniform type in \subref{fig:TypeUnifSubFig} and a binomial type in \subref{fig:TypeBinomSubFig}. Each data point was produced via an average over $40$ independent experiments. The black line in each plot represents the theoretical upper bound of $(1/\sqrt{m})$.}
\end{figure}

To empirically illustrate the expected $L_2$-error result of Lemma~\ref{lem:TypeProps}, we conducted simulations that are summarized in Figure~\ref{fig:TypeEvaluationExample}.
In the simulations, we randomly sampled synthetic sequences of two different types (uniform and binomial) and measured the expected $L_2$-error of the sampled type estimate. By varying the parameters $n$ (length of the sequences) and $m$ (number of samples), we can see that the expected $L_2$-error of the estimated type depends only on $m$ and is independent of $n$. The black line in each plot represents the $(1/\sqrt{m})$ theoretical upper bound.

\subsection{Obtaining the Function Estimate from the Joint Type} \label{sec:FuncEst}

The random sampling approach can be used to estimate any normalized
sum-type function by expressing it as a function of the joint type as
follows:
\begin{eqnarray*}
f_i(\mathbf{x}^k) &=& \frac{1}{n} \sum_{l=1}^n \phi_i(x_{1,l},\ldots,x_{k,l}) \\
&=& \frac{1}{n} \sum_{x^k} \phi_i(x^k) N(x^k) \\
&=& \sum_{x^k} \phi_i(x^k) P_{\mathbf{x}^k}(x^k).
\end{eqnarray*}
Let an estimate of $f_i(\mathbf{x}^k)$ based on only the sampled
sequences $(\mathbf{x}^k)_I$ be given by
\begin{eqnarray}
\hat{F}_i(\mathbf{x}^k) &:=& \frac{1}{m} \sum_{l \in L} \phi_i(x_{1,l},\ldots,x_{k,l}) \label{eqn:FuncEst} \\
&=& \frac{1}{m} \sum_{x^k} \phi_i(x^k) M(x^k) \label{eqn:AltFuncEst} \\
&=& \sum_{x^k} \phi_i(x^k) \hat{P}_{\mathbf{x}^k}(x^k). \label{eqn:FuncTypeUnified}
\end{eqnarray}
\begin{lem} \label{lem:FuncEstProps}
For any finite-alphabet sequences $(\mathbf{x}_1,\ldots,\mathbf{x}_k)
\in \mathcal{X}_1^n \times \ldots \times \mathcal{X}_k^n$, and
sampling parameter $m \in \{1, \ldots, n\}$, the expected absolute
error of the randomly sampled function estimate given by
(\ref{eqn:FuncEst}) is bounded by
\begin{eqnarray*}
E\left[\big|\hat{F}_i(\mathbf{x}^k) - f_i(\mathbf{x}^k)\big|\right] &\leq& \frac{\|\phi_i\|_2}{\sqrt{m}},
\end{eqnarray*}
where
\begin{eqnarray*}
\big\| \phi_i \big\|_2 := \sqrt{\sum_{x^k} |\phi_i(x^k)|^2}.
\end{eqnarray*}
\end{lem}

\IEEEproof
See Appendix~\ref{app:FuncEstProps} for the detailed proof.
\endproof

\begin{figure}[!htb]
\centering
\includegraphics[width=3.4in]{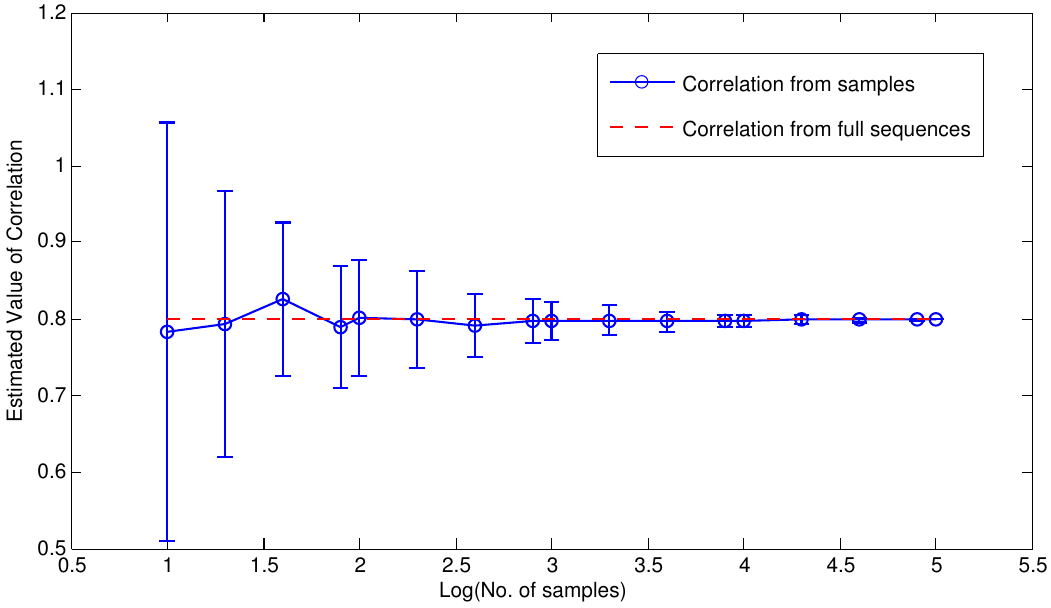}
\caption{Empirical correlation between sequences $\mathbf{x}$ and
  $\mathbf{y}$ estimated by random sampling.  The expected absolute
  error between the true and estimated correlations is inversely
  proportional to the square root of the sampling parameter and
  independent of the sequence length. Each data point corresponds to the expected value of the function over 100 realizations of sampling followed by function evaluation over the samples.}
\label{fig:CorrelationEvaluationExample}
\end{figure}

Thus, an estimate formed from random samples would satisfy the
error claims of Theorem~\ref{thm:MainThm}. To illustrate the
evaluation of a normalized sum-type function by randomized
sampling, suppose that $\mathbf{x} \in \{0,1\}^n$ and $\mathbf{y}
\in \{0,1\}^n$ are two scalar sequences and we are interested in
estimating their empirical correlation $\frac{1}{n}\sum_{l=1}^n x_l
y_l$. The sequences $\mathbf{x}$ and $\mathbf{y}$ represent two
separate databases. For example, $\mathbf{x}$ may represent the
presence or absence of a specific genetic marker in the DNA of a large
number of tested individuals, and $\mathbf{y}$ may represent the
incidence or non-incidence of a certain disease in those
individuals. A researcher may be interested in evaluating the
correlation between $\mathbf{x}$ and $\mathbf{y}$. Based on the
results of this section, the researcher can estimate the correlation
by randomly sampling $\mathbf{x}$ and
$\mathbf{y}$. Figure~\ref{fig:CorrelationEvaluationExample} plots the
correlation estimate as a function of the sampling parameter. The
error bars indicate the accuracy of the estimate, in terms of the
expected absolute error between the true and estimated correlation,
where the expectation is taken over $100$ independent experiments
each using a different realization of randomized sampling. As a
sanity check, the dashed line plots the true correlation computed
using full versions of the two sequences.

Our next Lemma establishes a large deviation bound for our function estimate.

\begin{lem} \label{lem:LargeDeviation}
For any finite-alphabet sequences $(\mathbf{x}_1,\ldots,\mathbf{x}_k)
\in \mathcal{X}_1^n \times \ldots \times \mathcal{X}_k^n$, and
sampling parameter $m \in \{1, \ldots, n\}$, the probability that
the absolute error of the randomly sampled function estimate given
by (\ref{eqn:FuncEst}) exceeds $\delta$ is bounded by
\begin{align*}
\Pr \Big[ \big|\hat{F}_i(\mathbf{x}^k) - f_i(\mathbf{x}^k)\big| \geq \delta \Big] \leq 2 \exp \left(\frac{-2 \delta^2 m}{\phi_i^{\mathrm{range}}} \right),
\end{align*}
where 
\[
\phi_i^{\mathrm{range}} := (\max_{x^k} \phi_i(x^k)) - (\min_{x^k} \phi_i(x^k)).
\]
\end{lem}

\IEEEproof
See Appendix~\ref{app:LargeDeviation} for the detailed proof.
\endproof

An immediate corollary of this result is that the type estimate
themselves will satisfy a similar large deviation bound.

\begin{cor}
For any finite-alphabet sequences $(\mathbf{x}_1,\ldots,\mathbf{x}_k)
\in \mathcal{X}_1^n \times \ldots \times \mathcal{X}_k^n$, sampling
parameter $m \in \{1, \ldots, n\}$, and $x^k \in \mathcal{X}_1 \times
\ldots \times \mathcal{X}_k$, the probability that the absolute error
of the randomly sampled type estimate given by (\ref{eqn:TypeEst})
exceeds $\delta$ is bounded by
\begin{align*}
\Pr \Big[ \big| \hat{P}_{\mathbf{x}^k} (x^k) - P_{\mathbf{x}^k} (x^k) \big| \geq \delta \Big] \leq 2 \exp(-2 \delta^2 m),
\end{align*}
\end{cor}

\IEEEproof For a given $x^k \in \mathcal{X}_1 \times \ldots \times
\mathcal{X}_k$, by (\ref{eqn:FuncTypeUnified}), the type estimate
$\hat{P}_{\mathbf{x}^k} (x^k)$ can be viewed as a specific function
estimate where $\phi_i(y^k)$ is equal to one if $y^k = x^k$ and zero
otherwise.  Thus, we can apply Lemma~\ref{lem:LargeDeviation}, with
$\phi_i^{\mathrm{range}} = 1$, to yield the result.  \endproof

So far we have shown that any normalized sum-type function of $k$
sequences of length $n$ can be accurately estimated from randomized
sampling with both vanishing distortion and vanishing communication
cost as $n \rightarrow \infty$. In the following section, we describe
how this computation can be performed in a distributed and
unconditionally private manner in a fully connected network, with
distortion and communication cost still going to zero as $n
\rightarrow \infty$.

\section{Protocols} \label{sec:Protocols}

In this section we provide three protocols that satisfy the properties claimed in Theorem~\ref{thm:MainThm}.
All of these protocols make use of unconditionally secure multi-party computation methods and random sampling to produce function estimates $\hat{F}_i(x^k)$ as given by (\ref{eqn:FuncEst}).
In Section~\ref{sec:KeyResults}, we showed that these function estimates are accurate as claimed in Theorem~\ref{thm:MainThm}.
In this section, we show that the protocols are unconditionally private and incur communication cost as claimed in Theorem~\ref{thm:MainThm}.
While these protocols are designed for fully connected networks, they can repurposed for certain types of partially connected newtorks as claimed by Theorem~\ref{thm:NetworkThm} and explained in its proof in Appendix~\ref{app:NetworkProof}
The first two protocols given in Section~\ref{sec:PolyProtocols} are for the general $k$-party case ($k \geq 3$) and utilize the secure multi-party computation methods of~\cite{BenOrGwW-ACM88-CTNCFTDC}.
In Section~\ref{sec:OTPprotocol}, we give a third protocol that only applies to a specific three-party scenario, however it is interesting since it only utilizes simple one-time pad masking techniques.

For all of our protocols, the common first step is for one of the parties to randomly choose the $m$ sampling locations $L \subset \{1,\ldots,n\}$, uniformly without replacement, and communicate them to the rest of the parties with $(k-1)m \log n$ bits.
From here, the specifics of the protocols differ, but they all require the parties to work with only the sampled sequences $(\mathbf{x}_1,\ldots,\mathbf{x}_k)_L$ and result in each party computing $\sum_{l \in L} \phi_i(x_{1,l},\ldots,x_{k,l}) = m\hat{F}_i(x^k)$ via finite field arithmetic.
Since the domain $\mathcal{X}_1 \times \ldots \times \mathcal{X}_k$ is finite, the range of each function $f_i$ is a finite subset of $\mathbb{Q}$.
Thus, with a sufficiently large finite field, $\mathcal{F}_m$, to prevent interger-arithmetic overflow, the computation of $\sum_{l \in L} \phi_i(x_{1,l},\ldots,x_{k,l})$ can be performed with finite field arithmetic in $\mathcal{F}_m$.
The computation can be expressed in a finite field of a size on the order of $O(m)$ since it is a sum of $m$ rational values from the finite image set of $\phi_i$.
The finite field representation of $\sum_{l \in L} \phi_i(x_{1,l},\ldots,x_{k,l})$ can then be converted back into a rational number and divided by $m$ to produce $\hat{F}_i(x^k)$.
To ease comparison, we have grouped the presentation of the steps of each protocol (via numbering) into four distinct phases: 1) sampling the sequences, 2) secret sharing of the inputs, 3) secure computation on the shares, and 4) revelation of the computed outputs.

All three protocols require $O(m \log |\mathcal{F}_m|)$ bits in addition to the $m (k-1) \log n$ bits required to transmit $L$.
Since the size of the finite field need only be on the order of $|\mathcal{F}_m| = O(m)$, the total number of bits needed is actually dominated by the transmission of $L$ and is on the order of $r = O(m \log n)$ as $m$ and $n$ scale.
In the following subsections, we will discuss and compare the specific communication costs of each protocol.

\subsection{Polynomial Secret-Sharing Based Protocols} \label{sec:PolyProtocols}

We present two protocols that employ the secure multi-party computation methods of~\cite{BenOrGwW-ACM88-CTNCFTDC}, which utilize the homomorphic properties of polynomial-based secret sharing~\cite{Shamir-ACM79-SecretSharing}.
These methods begin with each party generating and distributing secret shares of its private input to all of the parties.
These shares are unconditionally secure and do not reveal any information unless collectively recombined.
The homomorphic properties of these shares allow that appropriate additive and multiplicative operations on the shares produce shares of the corresponding additive and multiplicative computations of the underlying inputs.
By exploiting this property, shares of the desired computations can be computed and then collectively recombined by the appropriate parties in order to perform secure computation.
Our two protocols create the same function estimates, but do so via the different expanded forms given in (\ref{eqn:FuncEst}) and (\ref{eqn:AltFuncEst}).

The first protocol {\bf PolyTF} (``Type First'') is based on the function estimate expression given by (\ref{eqn:AltFuncEst}).
The parties first interactively compute homomorphic secret shares of the partial frequency function $M$.
By exploiting the additive form of function estimates based on $M$ and the homomorphic properties of the shares, the parties can combine the shares to compute the desired function estimates.
Computation of $M(x^k)$ can be performed using the following expansion
\begin{equation} \label{eqn:LShareComp}
M(x_1,\ldots,x_k) = \sum_{l \in L} \prod_{i \in \{1,\ldots,k \}} \mathbf{1}_{\{x_{i,l}\}}(x_i),
\end{equation}
where $\mathbf{1}_{\{x_{i,l}\}}(x_i)$ is the indicator function equal to one if $x_i = x_{i,l}$ and zero otherwise.
The {\bf PolyTF} protocol is outlined by the following steps:

\begin{enumerate}
\item The first party randomly chooses $m$ sampling locations $L \subset \{1,\ldots,n\}$, uniformly without replacement and communicates the index set $L$ to the remaining $k-1$ parties using $m (k-1) \log n$ bits.
\item For each $l \in L$ and $i \in \{1, \ldots, k\}$, party $i$ creates homomorphic secret shares of the $|\mathcal{X}_i|$-length vector representing the indicator function $\mathbf{1}_{\{x_{i,l}\}}(\cdot)$, and sends a share to each party. This uses a total of $m(k-1)(|\mathcal{X}_1| + \ldots + |\mathcal{X}_k|) \log |\mathcal{F}_m|$ bits across all parties.
\item[3a)] For each $x^k \in \mathcal{X}_1 \times \ldots \times \mathcal{X}_k$ and $i \in \{1, \ldots, k\}$, party $i$ produces a share of $M(x^k)$ via secure additive and multiplicative combinations of the shares of the indicator function vectors according to (\ref{eqn:LShareComp}). This step also requires communication since each multiplication of shares requires degree reduction and randomization of shares (see~\cite{BenOrGwW-ACM88-CTNCFTDC} for details). This step uses a total of $m(k-1)(|\mathcal{X}_1| \cdot \ldots \cdot |\mathcal{X}_k|)$ multiplications and each multiplication requires a total of $k (k-1) \log |\mathcal{F}_m|$ bits to be sent for randomization after degree reduction. Thus, this step requires a total of $mk(k-1)^2(|\mathcal{X}_1| \cdot \ldots \cdot |\mathcal{X}_k|) \log |\mathcal{F}_m|$ bits of communication.
\item[3b)] For each $i,j \in \{1, \ldots, k\}$, party $i$ computes a share of the function estimate $\hat{F}_j$ via
\[
\hat{F}_j[i] = \sum_{x^k} \phi_j(x^k) M_i(x^k),
\]
exploiting the additive form of the function estimate given by (\ref{eqn:AltFuncEst}) and the homomorphic properties of the shares.
\item[4)] For each $i \in \{1, \ldots, k\}$, the shares $\hat{F}_i[1], \ldots, \hat{F}_i[k]$ are sent to party $i$, who can recover $m \hat{F}_i(x_k)$, and hence the desired function estimate. This step uses $k(k-1) \log |\mathcal{F}_m|$ bits.
\end{enumerate}

The complexity of this protocol lies mainly in steps 2 and 3a, which create the shares of $M(x^k)$.
However, note that the complexity of generating these shares is independent of the desired function computations, and step 3b consists of only local computation, requiring no communication.
Thus, the overall communication requirements of this protocol are independent of the underlying functions to be computed, $f_1, \ldots, f_k$.
In addition to the $m (k-1) \log n$ bits needed to distribute the sampling set $L$, this protocol requires an additional $m(k-1) \big( (|\mathcal{X}_1| + \ldots + |\mathcal{X}_k|) + k(k-1)(|\mathcal{X}_1| \cdot \ldots \cdot |\mathcal{X}_k|) \big) \log |\mathcal{F}_m| + k(k-1) \log |\mathcal{F}_m|$ bits, which is of the order of $O(m \log |\mathcal{F}_m|)$.

The second protocol {\bf PolyDS} (``Direct Sum'') takes the direct approach of computing 
the sum given in (\ref{eqn:FuncEst}). The parties first securely compute shares of 
$\phi_j(x_{1,l},\ldots,x_{k,l})$ for each $j \in \{1, \ldots, k\}$ and $l \in L$.
These shares can then be additively combined to produce shares of the function estimates.
The {\bf PolyDS} protocol is outlined by the following steps:

\begin{enumerate}
\item The first party randomly chooses $m$ sampling locations $L \subset \{1,\ldots,n\}$, uniformly without replacement and communicates the index set $L$ to the remaining $k-1$ parties using $m (k-1) \log n$ bits.
\item For each $l \in L$ and $i \in \{1, \ldots, k\}$, party $i$ creates homomorphic secret shares of $x_{i,l}$, and sends a share to each party. This uses a total of $mk(k-1) \log |\mathcal{F}_m|$ bits across all parties.
\item[3a)] Using the secure computation methods of~\cite{BenOrGwW-ACM88-CTNCFTDC}, each party $i \in \{1, \ldots, k\}$ securely obtains a share of $\phi_j(x_{1,l},\ldots,x_{k,l})$, for each $j \in \{1, \ldots, k\}$ and $l \in L$, denoted by $\phi_j(x_{1,l},\ldots,x_{k,l})[i]$.
\item[3b)] For each $i,j \in \{1, \ldots, k\}$, party $i$ additively combines it shares to produce a share of the function estimate $\hat{F}_j$ via
\[
\hat{F}_j[i] = \sum_{l \in L} \phi_j(x_{1,l},\ldots,x_{k,l})[i],
\]
exploiting the additive form of the function estimate given by (\ref{eqn:FuncEst}) and the homomorphic properties of the shares.
\item[4)] For each $i \in \{1, \ldots, k\}$, the shares $\hat{F}_i[1], \ldots, \hat{F}_i[k]$ are sent to party $i$, who can recover $m \hat{F}_i(x_k)$, and hence the desired function estimate. This step uses $k(k-1) \log |\mathcal{F}_m|$ bits.
\end{enumerate}

The complexity of this protocol mainly lies in the step 3a, consisting of computing the shares of $\phi_j(x_{1,l},\ldots,x_{k,l})$.
The actual details of this step depend on the structures of $\phi_1, \ldots, \phi_k$ and how they can be represented by a multi-variate polynomial over a finite field, which, in principle, is always feasible by interpolation but could possibly result in a very high-degree polynomial.
Securely computing these functions thus boils down to securely performing the arithmetic operations of addition and multiplication that make up their respective polynomial representations.
While each addition can be performed using only localized computation, each multiplication requires additional transmissions amongst the parties to perform degree reduction and re-randomization on the shares (see~\cite{BenOrGwW-ACM88-CTNCFTDC} for details).
For each multiplication in a particular function $\phi_i$, $mk(k-1) \log |\mathcal{F}_m|$ bits of communication are required between all parties and across all samples.
Thus, letting $\psi_i$ denote the number of multiplications needed to compute $\phi_i$, the total number of bits required for step 3b is $mk(k-1) (\psi_1 + \ldots + \psi_k) \log |\mathcal{F}_m|$.
In addition to the $m (k-1) \log n$ bits needed to distribute the sampling set $L$, this protocol requires an additional $(m(\psi_1 + \ldots + \psi_k + 1) + 1) k(k-1) \log |\mathcal{F}_m|$ bits, which is of the order of $O(m \log |\mathcal{F}_m|)$.

\subsection{One-Time Pad Protocol for Three-Parties} \label{sec:OTPprotocol}

\begin{figure}[thpb]
\centering
\includegraphics[width=2.3in]{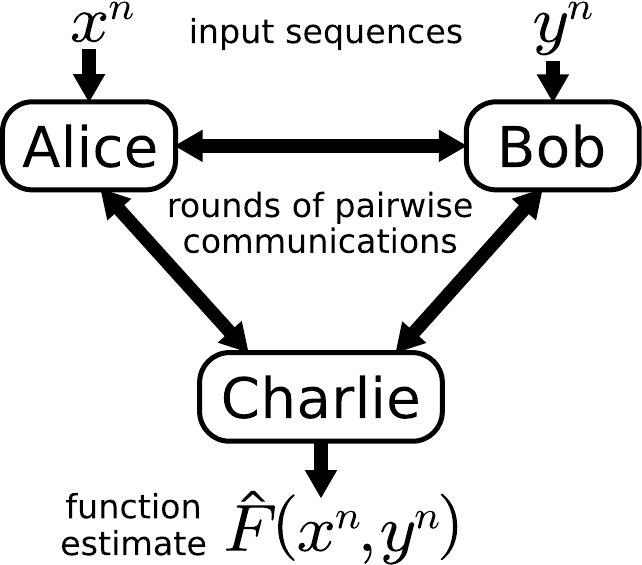}
\caption{Alice and Bob have deterministic sequences $x^n$ and $y^n$. Charlie produces an estimate of a function of the sequences, $\hat{F}(x^n,y^n)$.}
\label{fig:ThreeParty}
\end{figure}

In this section, we present a protocol for a specific three-party scenario, that is a simplified special case of our general problem setup.
Although the previously presented protocols could be applied to this scenario, the protocol that we discuss in this section is interesting since it leverages the simple techniques of one-time pad encryption and additive shares.
For clarity and ease of exposition, we present this three-party scenario with the parties named as Alice, Bob, and Charlie and use the following notation. 
Alice and Bob each have a sequence of $n$ symbols, denoted respectively by $x^n := (x_1,\ldots,x_n) \in \mathcal{X}^n$ and $y^n := (y_1,\ldots,y_n) \in \mathcal{Y}^n$, where $\mathcal{X}$ and $\mathcal{Y}$ are finite alphabets.
Charlie wishes to compute a normalized sum-type function of Alice and Bob's sequences, given by
\[
f(x^n,y^n) = \frac{1}{n} \sum_{l=1}^n \phi(x_l,y_l).
\]
This scenario is illustrated in Figure~\ref{fig:ThreeParty}.

For this three-party scenario, we propose the {\bf OTP} protocol, which leverages a homomorphic property of one-time pad encryption.
Alice and Bob respectively send their sampled sequences $(x_l)_{l \in L}$ and $(y_l)_{l \in L}$, masked (encrypted) with one-time pads, to Charlie.
From these encrypted sequences, Charlie computes and returns to Alice and Bob encrypted additive shares of the partial frequency function $M$.
Alice and Bob decrypt their respective messages from Charlie to obtain the additive shares of $M$, from which they derive additive shares of the function estimate that are returned to be recombined by Charlie. 
This technique of first computing additive shares of $M$, as an intermediate step, takes advantage of the function estimate expansion given by (\ref{eqn:AltFuncEst}).
The transmissions required by and the complexity of implementing this protocol are independent of the complexity of $\phi$ (except indirectly through the necessary size of $\mathcal{F}_m$).

The steps of the {\bf OTP} protocol are as follows:

\begin{enumerate}
\item  Alice randomly chooses $m$ sampling locations $L \subset \{1,\ldots,n\}$,
uniformly without replacement. She sends the set of sampling locations $L$ to Bob using $m \log n$ bits.
\item[2a)] Alice also generates two one-time pads $(\alpha_l)_{l \in L}$ and $(\beta_l)_{l \in L}$, by choosing $\alpha_l \sim \text{iid Unif}(\{0,\ldots,|\mathcal{X}| - 1\})$ and $\beta_l \sim \text{iid Unif}(\{0,\ldots,|\mathcal{Y}| - 1\})$.
She sends the two one-time pads to Bob using $m(\log|\mathcal{X}| + \log|\mathcal{Y}|)$ bits.
\item[2b)] Alice applies the pad $(\alpha_l)_{l \in L}$ to $(x_l)_{l \in L}$, producing the encrypted 
sequence $(\overline{x}_l)_{l \in L}$, by setting $\overline{x}_l = \oplus_{\alpha_l}(x_l)$, where $\oplus_{\alpha_l}(x_l)$ is a circular shift of the value of $x_l$ over an arbitrary ordering 
of $\mathcal{X}$ by $\alpha_l$ positions.
\item[2c)] Similarly, Bob applies the pad $(\beta_l)_{l \in L}$ to $(y_l)_{l \in L}$ to produce $(\overline{y}_l)_{l \in L}$ by setting $\overline{y}_l = \oplus_{\beta_l}(y_l)$.
\item[2d)] Alice and Bob respectively send $(\overline{x}_l)_{l \in L}$ and $(\overline{y}_l)_{l \in L}$ to Charlie, using a total of $m (\log |\mathcal{X}| + \log |\mathcal{Y}|)$ bits. As Charlie does not
know the one-time pads, he cannot recover Alice's and Bob's original samples from
the encryptions $(\overline{x}_l)_{l \in L}$ and $(\overline{y}_l)_{l \in L}$.
\item[3a)] For each $l \in L$, Charlie produces $Q_l$, an $|\mathcal{X}| \times |\mathcal{Y}|$ matrix indexed by $(x,y) \in \mathcal{X} \times \mathcal{Y}$, where $Q_l(x,y) = \mathbf{1}_{\{\overline{x}_l,\overline{y}_l\}}(x,y)$, which is the indicator function equal to one if $(\overline{x}_l,\overline{y}_l) = (x,y)$ and zero otherwise.
\item[3b)] Charlie splits each $Q_l$ into additive shares, by first independently choosing, across all $(l,x,y) \in L \times \mathcal{X} \times \mathcal{Y}$, $Q_{A,l}(x,y) \sim \text{iid Unif}(\mathcal{F}_m)$, then computing $Q_{B,l} = Q_l - Q_{A,l}$. 
\item[3c)] Charlie sends the matrices $(Q_{A,l})_{l \in L}$ to Alice and $(Q_{B,l})_{l \in L}$ to Bob, using a total of $2 m |\mathcal{X}| |\mathcal{Y}| \log |\mathcal{F}_m|$ bits. This splitting into additive
secret shares prevents Alice and Bob from finding out about each others sequences.
\item[3d)] Alice and Bob separately decrypt $(Q_{A,l})_{l \in L}$ and $(Q_{B,l})_{l \in L}$ to compute additive shares of $M$, via
\begin{eqnarray*}
M_A(x,y) = \sum_{l \in L} Q_{A,l}(\oplus_{\alpha_l}(x),\oplus_{\beta_l}(y)), \\
M_B(x,y) = \sum_{l \in L} Q_{B,l}(\oplus_{\alpha_l}(x),\oplus_{\beta_l}(y)).
\end{eqnarray*}
\item[3e)] Alice and Bob separately compute additive shares of the function computation, via
\begin{eqnarray*}
F_A = \sum_{x,y} \phi(x,y) M_{A}(x,y), \\
F_B = \sum_{x,y} \phi(x,y) M_{B}(x,y).
\end{eqnarray*}
\item[4a)] Alice independently generates random salt $Z$ uniformly over $\mathcal{F}_m$, which she sends to Bob using $\log |\mathcal{F}_m|$ bits. 
The random salt is used to maintain security by statistically decorrelating Alice and Bob's function estimate shares from information already held by Charlie.
\item[4b)] Alice and Bob send $F_A + Z$ and $F_B - Z$ to Charlie using a total of $2 \log |\mathcal{F}_m|$ bits. Note that $F_A + F_B = m \hat{F}(x^n,y^n)$ because of the definition of $Q_l(x,y)$. Thus, Charlie can produce $\hat{F}(x^n,y^n)$.
\end{enumerate}

Thus, in addition to the $m \log n$ bits needed to convey the sampling set $L$ from Alice to Bob, the {\bf OTP} protocol requires an additional
$2 m(\log |\mathcal{X}| + \log |\mathcal{Y}| + |\mathcal{X}| |\mathcal{Y}| \log |\mathcal{F}_m|) + 3 \log |\mathcal{F}_m|$ bits, which is of the order of $O(m \log |\mathcal{F}_m|)$.

\subsection{Comparison of Protocols}

All of our protocols are unconditionally private and produce the same function estimates $\hat{F}_i(x_k)$ while requiring $m (k-1) \log n + O(m \log |\mathcal{F}_m|)$ bits.
Their subtle performance differences are in the specific constants of the $O(m \log |\mathcal{F}_m|)$ term.
The relative performance of the protocols varies based on the specific functions that are required to be computed.
Table~\ref{tab:ProtComp} summarizes the comparison of our three protocols, with the communication costs broken down across the four phases of: 1) sampling the sequences, 2) secret sharing of the inputs, 3) secure computation on the shares, and 4) revelation of the computed outputs.
For each phase, the first row represents the general costs, while the second row represents the costs when specialized to the specific three-party setup of Section~\ref{sec:OTPprotocol}, allowing for the comparison of all three protocols in this particular scenario, since the {\bf OTP} protocol is only applicable to the specific three-party scenario described by Section~\ref{sec:OTPprotocol}

For functions where the $\phi_i$ functions can be represented as simple polynomials, the {\bf PolyDS} protocol is the simplest and most efficient, possibly using only $(m+1)k(k-1) \log |\mathcal{F}_m|$ bits (in addition to the $m (k-1) \log n$ needed to transmit $L$) when the computation of the $\phi_i$ functions require no multiplications.
However, for functions where $\phi_i$ are more complicated (e.g., containing absolute values or thresholding), requiring a high-degree polynomial representation, the complexity and the bits needed for protocol {\bf PolyDS} increase.
For such functions it may be better to use the {\bf PolyTF} protocol, which computes, as an intermediate step, homomorphic shares of the partial frequency function $M$, which is then used to generate shares of any function estimates at a fixed additional communication cost.
The complexity and bits required by the {\bf PolyTF} and {\bf OTP} protocols are not affected by the complexity of the $\phi_i$ functions (except indirectly through the necessary size of $\mathcal{F}_m$), and hence are more efficient than the {\bf PolyDS} protocol for very complex $\phi_i$ functions.
The {\bf OTP} protocol is only applicable to the specific three-party scenario described by Section~\ref{sec:OTPprotocol}, however it is potentially more efficient than both the {\bf PolyTF} and {\bf PolyDS} protocols, and also of interest since it demonstrates how the simple techniques of one-time pad encryption and additive shares are sufficient to construct a secure computation protocol for this specific three-party scenario.

\begin{table*}[!t]
\center
\caption{Comparison of the communication costs of our secure computation protocols \label{tab:ProtComp}}
\renewcommand{\arraystretch}{1.5}
\begin{tabular}{| l || c | c | c |}
\hline
Protocol & {\bf PolyTF} & {\bf PolyDS} & {\bf OTP}
\\ \hline \hline
Sampling & \multicolumn{2}{|c|}{$m(k-1) \log(n)$} & N/A
\\ \hline 
(\emph{3-party}) & \multicolumn{3}{|c|}{$m \log(n)$}
\\ \hline \hline
Sharing & $m (k-1)(\sum_i |\mathcal{X}_i|) \log|\mathcal{F}_m|$
        & $m k (k-1)\log|\mathcal{F}_m|$
        & N/A
\\ \hline 
(\emph{3-party}) & $2m (|\mathcal{X}| + |\mathcal{Y}|) \log|\mathcal{F}_m|$
                 & $4m \log |\mathcal{F}_m|$
                 & $2m (\log |\mathcal{X}| + \log |\mathcal{Y}|)$
\\ \hline \hline
Computation & $m k (k-1)^2 (\prod_i |\mathcal{X}_i|) \log|\mathcal{F}_m|$
            & $m k (k-1) (\sum_i \psi_i) \log |\mathcal{F}_m|$
            & N/A
\\ \hline 
(\emph{3-party}) & $6m (|\mathcal{X}| |\mathcal{Y}|) \log|\mathcal{F}_m|$
                 & $6m \psi \log |\mathcal{F}_m|$
                 & $2m |\mathcal{X}| |\mathcal{Y}| \log|\mathcal{F}_m|$
\\ \hline \hline
Revelation & \multicolumn{2}{|c|}{$k(k-1) \log|\mathcal{F}_m|$} & N/A
\\ \hline 
(\emph{3-party}) & \multicolumn{2}{|c|}{$2 \log|\mathcal{F}_m|$}
                 & $3 \log|\mathcal{F}_m|$
\\ \hline \hline
{\bf Total} & \multicolumn{3}{|c|}{$O(m \log n)$}
\\ \hline
\end{tabular}
\end{table*}

\section{Privacy of the Approximation} \label{sec:Privacy}

The protocols that we propose in Section~\ref{sec:Protocols} compute sampled approximations of the desired functions (according to the technique presented and analyzed in Section~\ref{sec:KeyResults}) along with revealing the sampling locations $L$ used to produce those estimates.
Viewing the protocol output as being the concatenation of the estimate with the sampling locations, these protocols satisfy the notion of unconditional protocol privacy as formulated in Section~\ref{sec:Formulation}, which requires that the protocols do not reveal anything more than these intended outputs.
In this section, we will explore the notion of {\em approximation privacy}, the second condition for privacy that we mentioned in Section~\ref{sec:Formulation}, which should require that the approximate computation not reveal any more information than must be inherently revealed by the exact computation.

The notion of {\em functional privacy} introduced by \cite[Definition~5]{FeigenbaumEtAl-06-SMCApprox}) is the most stringent form of approximation privacy.
An approximate computation is said to have perfect functional privacy if it can be
simulated (in a statistically indistinguishable sense) using only the
results of the exact computation without access to the original
inputs. Thus, perfect functional privacy demands that the distribution
of the approximate computation be only parameterized by the value(s)
of the exact computation, that is,
\[
P_{\hat{F}}(\hat{f};\mathbf{x}^k) = P_{\hat{F}}(\hat{f};f(\mathbf{x}^k)).
\]
If one replaces the deterministic input sequences with random
variables of any distribution, the above condition for perfect
functional privacy would imply the following Markov chain,
\[
\hat{F}(\mathbf{X}^k) - f(\mathbf{X}^k) - \mathbf{X}^k.
\]

Functional privacy, while providing a very strong notion of approximation privacy, can be rather restrictive for practical applications.
In particular, requiring functional privacy would rule out any random sampling-based approximations for general sum-type functions when $\phi(\cdot)$ takes more than two values, even if the sampling locations are concealed.
A sampling-based approximation for a normalized sum-type functions when the sample-wise function $\phi(\cdot)$ is binary-valued (e.g., joint/marginal types or normalized Hamming weight/distance) would be functionally private (provided that the sampling locations are not revealed).
This is because since $\phi(\cdot)$ is binary-valued, the exact computation $f(\mathbf{x}^k)$ would reveal the number of locations $l \in \{1, \ldots, n\}$ where $\phi(x_{1,l}, \ldots, x_{k,l})$ equals one of the binary values as opposed to the other, and hence a randomly sampled approximation could be simulated from knowledge of $f(\mathbf{x}^k)$ alone.
However, the sampling-based approximation is not functionally private for general normalized sum-type functions when $\phi(\cdot)$ may take more than two values.
To illustrate with a concrete example, consider a ternary-valued $\phi(\cdot)$ with range $\{0, 1, 2\}$, and potential input sequences $\mathbf{x}^k$ and $\mathbf{y}^k$ for which
\begin{align*}
( \phi(x[1]), \ldots, \phi(x[n]) ) &= (1, 1, \ldots, 1) , \\
( \phi(y[1]), \ldots, \phi(y[n]) ) &=
\begin{cases}
(0, 2, 0, 2, \ldots, 0, 2), & \text{$n$ even} \\
(0, 2, 0, 2, \ldots, 0, 2, 1), & \text{$n$ odd}
\end{cases}
\end{align*}
where $x[l] := x_{1,l}, \ldots, x_{k,l}$ and $y[l] := y_{1,l}, \ldots,
y_{k,l}$.
The exact computations for both input sequences are equal, that is, $f(\mathbf{x}^k) = f(\mathbf{y}^k) = 1$, however, the distributions of randomly sampled approximations for inputs
$\mathbf{x}^k$ and $\mathbf{y}^k$ are different:
$\hat{F}(\mathbf{x}^k)$ is equal to $1$ with probability one, while
$\hat{F}(\mathbf{y}^k)$ is hypergeometrically distributed.
This shows that the randomly sampled estimates cannot be perfectly simulated from
only the exact function computation without access to the original input sequences.
In general, when the sample-wise function $\phi(\cdot)$ can take more than two different values, there exist many functions for which the distribution of the randomly sampled estimate can be used to statistically distinguish between different sets of input sequences that produce identical outputs when the function is computed exactly.

The difficulty of practically achieving functional privacy motivates us to argue for a weaker notion of approximation privacy applicable to the asymptotic regime that we are interested in.
The approximation privacy of our sampling-based estimate is affected by the choice of the sampling parameter $m$ in relation to the overall sequence length $n$.
For example, in one extreme, when $m$ is equal to one (or similarly small and fixed), the subsampling estimate would be inaccurate and provide substantially different information than the exact computation (namely, the sampled computation over a small subset of the data, which in general is not likely to be representative of the exact computation).
On the other extreme, when $m$ is equal to $n$, that is the entire sequence is sampled, the computation would be exact and the concern of approximation privacy would become irrelevant.
However, the particular asymptotic scenario that we are interested in is when the number of samples $m$ increases to infinity, but the ratio of samples to total sequence length $m/n$ decreases to zero, that is, the sequence length $n$ grows faster than the sampling parameter $m$.
This is the asymptotic where both the distortion and communication cost become vanishing (as shown by Corollary~\ref{cor:vanishing}), and where we will argue that the approximate computation is essentially as private as the exact computation.

Consider a hypothetical scheme that produces an output consisting of the exact computation $f(\mathbf{x}^k)$ along with $m$ locations $L_1, \ldots, L_m$ that are independently chosen uniformly without replacement from $\{1, \ldots, n\}$.
Clearly, this output would not reveal any more information about the inputs than the exact computation alone, and this hypothetical scheme would be perfectly functionally private.
Thus, there is no loss in the approximation privacy by appending a list of random independent locations to the exact computation, producing the output $(f(\mathbf{x}^k), L_1, \ldots, L_m)$.
The output of our schemes swaps the exact computation with a sampling-based estimate that is dependent on the random locations, $(\hat{F}(\mathbf{x}^k), L_1, \ldots, L_m)$, specifically the normalized function computed only over these locations, which may reveal information that is substantially different than the exact computation.
The concern of approximation privacy for our schemes thus boils down to whether this output reveals substantially different information than the exact computation appended to random locations.

In our asymptotic of interest (with both $n$ and $m$ growing, but allowing $m/n$ to become vanishing), we have established that the approximation $\hat{F}(\mathbf{x}^k)$ converges to the exact computation $f(\mathbf{x}^k)$, for any input sequences $\mathbf{x}^k$, in terms of both mean error (by Corollary~\ref{cor:vanishing}) and probability (following from Lemma~\ref{lem:LargeDeviation}).
We propose that the property of convergence in probability provides a reasonable level of approximation privacy for schemes in this asymptotic.
Concretely, the level of approximation privacy is given by a pair parameters $\delta, \epsilon > 0$, which is said to be satisfied if, for all sequences $\mathbf{x}^k$,
\[
\Pr \Big[ \big| \hat{F}(\mathbf{x}^k) - f(\mathbf{x}^k) \big| \geq \delta \Big] \leq \epsilon,
\]
and where the objective is to minimize both $\delta$ and $\epsilon$.
Using this convergence property to define the condition for approximation privacy can be interpreted as: for a party to gain any significant information from the approximation beyond that obtained from the exact computation, they would have to encounter the event that the random sampling chooses locations for which the approximate computation is significantly different than the exact computation.
Lemma~\ref{lem:LargeDeviation} implies that the sampling approximation can be made to fall within an arbitrarily small neighborhood ($\delta \rightarrow 0$) of the exact computation with arbitrarily high probability ($\epsilon \rightarrow 0$), in the asymptotic of growing sample size $m$ (even if the sampling ratio $m/n$ diminishes to zero), and thus the schemes can achieve an arbitrary level of approximation privacy.
This consequence of the lemma can be interpreted as follows: despite the scheme revealing the sampled approximation along with the sampling locations rather than the exact computation over the full sequences, the approximation becomes arbitrarily representative of (and hence does not reveal information substantially different from) the exact computation, in the asymptotic of interest.

A natural remaining question is whether the schemes could be further improved by concealing the sampling locations, while still keeping the communication cost on the same vanishing order.
Concealment of the sampling conditions could be achieved with the general secure multi-party computation techniques of~\cite{BenOrGwW-ACM88-CTNCFTDC}, however, these would require total bits exchanged for communications to be on the order of the full sequence length $n$, defeating the purpose of communication cost savings via sampling.
To the best of our knowledge, all known approaches for performing sampling allowing for the computation of the sampled approximation of general (nontrivial) functions, while also concealing the sampling locations and providing unconditional privacy, would require total bits exchanged for communications to be at least $O(n)$.
We conjecture that any such scheme may inherently require total bits exchanged to be on the order $n$, and hence that vanishing communication cost cannot be achieved for a sampling-based approach that conceals sampling locations.
Further examination of this conjecture is left for future work.

\section{Concluding Remarks}

This paper has introduced a distortion-theoretic approach for secure multi-party computation with unconditional privacy.
By generalizing the dimensionality reduction via sampling technique of~\cite{AhlswedeZhang-LNCS06-EstDist}, we have constructed protocols that securely compute any normalized sum-type function with arbitrarily high accuracy and vanishing communication cost.
This result is particularly relevant in secure statistical analysis of distributed databases, since many empirical statistical measures can be represented as normalized sum-type functions and overwhelming data sizes can be overcome by providing an efficient and accurate approximate computation.
The technique of randomized sampling allowed us to overcome the worst-case distortion criterion, yielding the result that for any sequences $(\mathbf{x}_1,\ldots,\mathbf{x}_k)$ of any arbitrary length $n$, the expected absolute error of the function estimate constructed from only $m$ random samples is inversely proportional to $\sqrt{m}$.

Future directions of work for this problem include exploring how these dimensionality reduction techniques can be applied to a more general class of functions.
One possible extension is to consider general functions of the joint type of the sequences (i.e., permutation-invariant functions).
However, challenges may arise for those functions that are sensitive to small errors in the joint type estimate, such as parity functions.
Another direction for future work is to formulate and extend the results to the two-party scenario, which requires external randomness.
In the two-party scenario ($k = 2$), in general, secure computation cannot be performed from scratch \cite{ChorKush-91-ZeroOneLaw}.
To extend these results to a two-party scenario, the randomized sampling technique could be paired with secure function computation techniques that utilize an oblivious transfer primitive~\cite{Kilian-ACM88-CryptoFromOT} or a binary erasure channel (see~\cite{WangIshwarISIT09}).
A notion of communication cost similar to~\cite{WangIshwarISIT09} could be defined by also counting the number of erasure channel uses or oblivious transfer primitive uses in addition to bits of error-free communication and dividing by $n$.
In these extensions, it would also be possible to prove similar results on achieving vanishing distortion and vanishing communication cost, while maintaining unconditional security.

\bibliographystyle{IEEEtran}
\bibliography{references}

\appendices
\renewcommand{\theequation}{\thesection.\arabic{equation}}
\setcounter{equation}{0}

\section{Proof of Theorem~\ref{thm:MainThm}}
\label{app:MainProof}

We will prove the theorem in two stages:
\begin{enumerate}
\item Show that the sampling-based estimate achieves vanishing distortion.
\item Demonstrate protocols that compute this estimate securely with vanishing communication cost.
\end{enumerate}

In Section~\ref{sec:KeyResults}, we analyze the approximation for normalized sum-type functions estimated from randomly sampling the sequences.
Lemma~\ref{lem:FuncEstProps} shows that this function estimate, given in (\ref{eqn:FuncEst}), meets the accuracy requirement of the theorem given by (\ref{eqn:ThmErrorBound}).
We leverage this result and construct protocols, given in Section~\ref{sec:Protocols}, that securely and exactly realize this function approximation.
All of our protocols securely compute this sampled estimate, using general secure computation techniques to provide security, while leveraging the sampling to achieve the claimed bit rate. 
In Section~\ref{sec:Protocols}, we determine more precise bit rates, which depend on function structure and choice of protocol.
However, all of the protocols achieve rates on the same order as claimed in the theorem.

\section{Proof of Theorem~\ref{thm:NetworkThm}}
\label{app:NetworkProof}

The protocols in Section~\ref{sec:Protocols} require a fully connected
network since messages must be sent between each pair of parties over
a private channel.  However, with a partially connected network, one
can simulate private channels routed through other parties in order to
simulate a fully connected network.  The secure transmission results
of \cite{DolevDQY-93-SecMsgTrans} establish that a message can be
privately transmitted from a source-vertex A to a destination-vertex B
using $t+1$ independent links between A and B provided that a passive
adversary can only eavesdrop on up to $t$ of those links.  This can be
done by using a secret sharing scheme to split the message into $t+1$
shares, of which all of them need to be known in order to recover the
message (see \cite{DolevDQY-93-SecMsgTrans} for details).  Menger's
theorem \cite{Harary-1994-GraphTheory} states that since $G$ has
minimum vertex cut greater than $t$, there exist at least $t+1$
vertex-independent paths between any pair of non-adjacent vertices.
Whenever a message needs to be sent between a pair of parties that are
not directly connected, we can use the techniques of
\cite{DolevDQY-93-SecMsgTrans} to privately transmit the message,
which is split into shares and routed over the $t+1$ non-overlapping
paths.  Since there are $t+1$ independent links, no coalition of size
$t$ parties can eavesdrop on the transmission, and hence a private
channel is effectively simulated.  The need to route the message over
multiple paths increases the total communication cost, however, since
$k$ is held fixed, the order with respect to $m$ and $n$ remains the
same, i.e., $O(m \log n)/n$.

\section{Proof of Lemma~\ref{lem:TypeProps}}
\label{app:TypeProps}

Observing that $M(x^k)$ has a hypergeometric distribution, the mean
and variance of the type estimate $\hat{P}_{\mathbf{x}^k}(x^k)$ can be
given by
\begin{eqnarray*}
E\big[\hat{P}_{\mathbf{x}^k}(x^k)\big] &=& \frac{E[M(x^k)]}{m} 
 = \frac{N(x^k)}{n} = P_{\mathbf{x}^k}(x^k), \\
\mathrm{Var}\big[\hat{P}_{\mathbf{x}^k}(x^k)\big] &=& \frac{\mathrm{Var}[M(x^k)]}{m^2} \\
&=& \frac{N(x^k)(n-N(x^k))(n-m)}{m n^2 (n-1)} \\
&\leq& \frac{N(x^k)}{m n} = \frac{P_{\mathbf{x}^k}(x^k)}{m}.
\end{eqnarray*}
Thus, the mean squared error summed across $x^k \in \mathcal{X}_1 \times \ldots \times \mathcal{X}_k$ is given by
\begin{eqnarray*}
\Sigma_{\mathrm{MSE}} &:=& E \left[ \sum_{x^k} \big|\hat{P}_{\mathbf{x}^k} (x^k) - P_{\mathbf{x}^k} (x^k) \big|^2 \right] \\
&=& \sum_{x^k} \mathrm{Var}\big[\hat{P}_{\mathbf{x}^k}(x^k)\big] \leq \frac{1}{m}.
\end{eqnarray*}
Continuing with Jensen's inequality yields a bound on the expected $L_2$ norm of the error,
\begin{eqnarray*}
E \left[ \big\|\hat{P}_{\mathbf{x}^k} - P_{\mathbf{x}^k} \big\|_2 \right] \leq \sqrt{\Sigma_{\mathrm{MSE}}} \leq \frac{1}{\sqrt{m}}.
\end{eqnarray*}

\section{Proof of Lemma~\ref{lem:FuncEstProps}}
\label{app:FuncEstProps}

The absolute error of the function estimate is bounded by
\begin{eqnarray*}
\big|\hat{F}_i(\mathbf{x}^k) - f_i(\mathbf{x}^k)\big| &=& \Big| \sum_{x^k} \phi_i(x^k) \big(\hat{P}_{\mathbf{x}^k}(x^k) - P_{\mathbf{x}^k}(x^k)\big) \Big| \\
&\leq& \big\| \phi_i \big\|_2 \cdot \big\|\hat{P}_{\mathbf{x}^k} - P_{\mathbf{x}^k} \big\|_2,
\end{eqnarray*}
due to the Cauchy-Schwartz inequality, where
\begin{eqnarray*}
\big\| \phi_i \big\|_2 = \sqrt{\sum_{x^k} |\phi_i(x^k)|^2}.
\end{eqnarray*}
Thus, the expected absolute error is bounded by
\begin{eqnarray*}
E\left[\big|\hat{F}_i(\mathbf{x}^k) - f_i(\mathbf{x}^k)\big|\right] &\leq& \frac{\|\phi_i\|_2}{\sqrt{m}}.
\end{eqnarray*}

\section{Proof of Lemma~\ref{lem:LargeDeviation}}
\label{app:LargeDeviation}

Our function estimate given by (\ref{eqn:FuncEst}) can be rewritten as follows,
\[
\hat{F}_i(\mathbf{x}^k) := \frac{1}{m} \sum_{l=1}^m Y_l,
\]
where
\[
Y_l := \phi_i(x_{1,L_l},\ldots,x_{k,L_l}),
\]
illustrating that the estimate is a normalized sum of random variables drawn uniformly without replacement from the set
\[
\left\{\phi_i(x_{1,l},\ldots,x_{k,l})\right\}_{l=1}^{n}.
\]
Since the empirical mean of the set is
\[
\frac{1}{n} \sum_{l=1}^n \phi_i(x_{1,l},\ldots,x_{k,l}) = f_i(\mathbf{x}^k),
\]
and each element is bounded in the range $[\min_{x^k} \phi_i(x^k), \max_{x^k} \phi_i(x^k)]$, we can apply Hoeffding's inequality \cite[Thm.~2]{Hoeffding-JASA63-ProbIneqForRandSums}, which is applicable to normalized sums of random variables drawn uniformly without replacement (see \cite[Sec.~6]{Hoeffding-JASA63-ProbIneqForRandSums}), to yield the desired result,
\[
\Pr \Big[ \big|\hat{F}_i(\mathbf{x}^k) - f_i(\mathbf{x}^k)\big| \geq \delta \Big] \leq 2 \exp \left( \frac{-2 \delta^2 m}{\phi_i^{\mathrm{range}}} \right).
\]

\end{document}